\documentclass[twocolumn,showpacs,showkeys,amsmath,amssymb]{revtex4}

\usepackage{graphicx}
\usepackage{dcolumn}
\usepackage{bm}
\usepackage{xspace}

\def\deg{$^\circ$}
\def\gcm2{g/cm$^2$}

\def\eref#1{(\ref{#1})\xspace}

\def\fref#1{Fig.\,\ref{#1}\xspace}
\def\ffref#1{Figs.\,\ref{#1}\xspace}

\def\rref#1{Ref.\,\cite{#1}\xspace}
\def\sref#1{Sect.\,\ref{#1}\xspace}
\def\and{\&\xspace}

\begin{document}

\preprint{APS/123-QED}

\title{A new method to measure the attenuation of hadrons in extensive air
showers}

\author{
W.D.~Apel$^1$, 
J.C.~Arteaga$^{2}$\footnote{now at Institute of Physics and Mathematics,
Universidad Michoacana, Morelia, Mexico},
F.~Badea$^{1}$,
K.~Bekk$^1$, 
M.~Bertaina$^3$,
J.~Bl\"umer$^{1,2}$,
H.~Bozdog$^1$,
I.M.~Brancus$^4$,
M.~Br\"uggemann$^5$,
P.~Buchholz$^5$,
E.~Cantoni$^{3,7}$,
A.~Chiavassa$^3$,
F.~Cossavella$^2$, 
K.~Daumiller$^1$, 
V.~de Souza$^2$\footnote{now at Universidade de S{\~a}o Paulo, Instituto de
Fisica de S{\~a}o Carlos, Brasil}, 
F.~Di~Pierro$^3$,
P.~Doll$^1$, 
R.~Engel$^1$,
J.~Engler$^1$, 
M.~Finger$^1$, 
D.~Fuhrmann$^6$,
P.L.~Ghia$^7$,
H.J.~Gils$^1$,
R.~Glasstetter$^6$, 
C.~Grupen$^5$,
A.~Haungs$^1$, 
D.~Heck$^1$, 
D.~Hildebrand$^{2}$\footnote{corresponding author, now at ETH Z\"urich, Switzerland, email: dorothee.hildebrand@phys.ethz.ch},
J.R.~H\"orandel$^{2}$\footnote{now at: Dept. of Astrophysics, Radboud
                               University Nijmegen, The Netherlands}, 
T.~Huege$^1$, 
P.G.~Isar$^1$, 
K.-H.~Kampert$^6$,
D.~Kang$^2$,
D.~Kickelbick$^5$,
H.O.~Klages$^1$, 
Y.~Kolotaev$^5$,
P.~{\L}uczak$^8$, 
H.J.~Mathes$^1$, 
H.J.~Mayer$^1$, 
J.~Milke$^1$, 
B.~Mitrica$^4$,
C.~Morello$^7$,
G.~Navarra$^3$,
S.~Nehls$^1$,
J.~Oehlschl\"ager$^1$, 
S.~Ostapchenko$^1$\footnote{now at Norwegian University, Trondheim, Norway},
S.~Over$^5$,
M.~Petcu$^4$, 
T.~Pierog$^1$, 
H.~Rebel$^1$, 
M.~Roth$^1$, 
H.~Schieler$^1$, 
F.~Schr\"oder$^1$,
O.~Sima$^9$, 
M.~St\"umpert$^2$, 
G.~Toma$^4$, 
G.C.~Trinchero$^7$,
H.~Ulrich$^1$,
J.~van~Buren$^1$,
W.~Walkowiak$^5$,
A.~Weindl$^1$,
J.~Wochele$^1$, 
M.~Wommer$^1$,
J.~Zabierowski$^8$
}
\affiliation{
$^{1}$ Institut\ f\"ur Kernphysik, Forschungszentrum Karlsruhe,
76021~Karlsruhe, Germany\\
$^{2}$ Institut f\"ur Experimentelle Kernphysik,
Universit\"at Karlsruhe, 76021 Karlsruhe, Germany,\\
$^{3}$ Dipartimento di Fisica Generale dell'Universit{\`a},
10125 Torino, Italy\\
$^{4}$ National Institute of Physics and Nuclear Engineering,
7690~Bucharest, Romania\\
$^{5}$ Fachbereich Physik, Universit\"at Siegen, 57068 Siegen, 
Germany \\
$^{6}$ Fachbereich Physik, Universit\"at Wuppertal, 42097
Wuppertal, Germany \\
$^{7}$ Istituto di Fisica dello Spazio Interplanetario, INAF, 
10133 Torino, Italy \\
$^{8}$ Soltan Institute for Nuclear Studies, 90950~Lodz, 
Poland\\
$^{9}$ Department of Physics, University of Bucharest, 
76900~Bucharest, Romania\\
}

\date{\today}

\begin{abstract}
Extensive air showers are generated through interactions of high-energy cosmic
rays impinging the Earth's atmosphere.  A new method is described to infer the
attenuation of hadrons in air showers.  The numbers of electrons and muons,
registered with the scintillator array of the KASCADE experiment are used to
estimate the energy of the shower inducing primary particle.  A large hadron
calorimeter is used to measure the hadronic energy reaching observation level.
The ratio of energy reaching ground level to the energy of the primary particle
is used to derive an attenuation length of hadrons in air showers.  In the
energy range from $10^6$~GeV to $3\cdot10^7$~GeV the attenuation length
obtained increases from 170~\gcm2 to 210~\gcm2.  The experimental results are
compared to predictions of simulations based on contemporary high energy
interaction models. 
\end{abstract}

\maketitle

\section{Introduction}\label{intro}

Since the earliest days of cosmic-ray investigations it has been realized that
these particles provide a unique possibility to study interactions at
high energies \cite{froman,rossigreisen}. Even today the energies of
cosmic rays exceed the energies achieved in man made accelerators by orders of
magnitude. Hence, in the literature many attempts are described to extract
properties of high-energy hadronic interactions from air showers induced by
cosmic rays in the atmosphere. Among the most interesting quantities is the
attenuation length of hadrons (e.g.\ \cite{handbphys1} p.\ 162), in theoretical
considerations closely connected to the inelastic cross section. 

In the present work, we use the energy absorbed in a material within a certain
atmospheric depth $X$ to define an attenuation length.  In this new approach we
use the number of electrons and muons, registered with a detector array to
estimate the energy of the shower inducing primary particle $E_0$, see
\eref{energyeq} below. The energy reaching the observation level in form of
hadrons $\sum E_H$ is measured with a hadron calorimeter. The fraction of
surviving energy in form of hadrons is defined as 
\begin{equation}\label{fractioneq}
 R=\frac{\sum E_H}{E_0} .
\end{equation}
The attenuation length $\lambda_E$ is then defined as
\begin{equation}\label{eatteq}
 \Sigma E_H=E_0\exp\left(-\frac{X}{\lambda_E}\right)  
\end{equation}
or 
\begin{equation}\label{ratioeq}
 R=\exp\left(-\frac{X}{\lambda_E}\right) .
\end{equation}

\ifnum 1=2

The interaction length $\lambda_{int}$ of protons in air is given as
\begin{equation}
 \lambda_{int}=\frac{A_{air}m_p}{\sigma_{p-air}} ,
\end{equation}
where $A_{air}=14.5$ is the average atomic weight of air, $m_p$ the proton
mass, and $\sigma_{p-air}$ the inelastic proton-air cross section.  The
experimentally accessible attenuation length is not only sensitive to the
proton-air interaction length, but also to the inelasticity of the
interactions.  
This dependence is frequently expressed by introducing a factor
$k$ \cite{block99,block00}: 
\begin{equation} \label{blockeq}
 \lambda=k\lambda_{int} .
\end{equation}
In practice, $k$ is also influenced by statistical fluctuations during the
shower development.
Measurements at the CERN SPS, where the complete showers of hadrons
with fixed energies up to 350~GeV \footnote{All energies in this article are
laboratory energies per particle.}  have been absorbed in a calorimeter,
indicate values $k\approx1.1$ \cite{kalocern}.

\fi

In the literature different definitions for the attenuation length in air
showers are introduced.  Frequently, the attenuation length is derived from
measurements of the electromagnetic shower component, e.g.\
\cite{kascadeabslength,hara,honda,eastopwq99,eastopwqprd,flyseyewq84,belovisvhecri}.
Investigating single hadrons the attenuation length is related to the
absorption of hadrons in the atmosphere \cite{yodh72}.  Pioneering work to
derive inelastic cross sections from the measurement of single hadrons has been
conducted by Yodh and colleagues \cite{yodh72,ellsworth}, later followed by the
prototype of the KASCADE calorimeter \cite{mielkesh}.

The new approach presented here is complementary to the different methods
described in the literature.  In contrast to methods using the electromagnetic
shower component, the present work focuses directly on measurements of hadrons
to derive an attenuation length for this shower component.  The values obtained
are not a priori comparable to other attenuation lengths since they are based
on different definitions.  It should be noted that the experimentally obtained
attenuation length is affected by statistical fluctuations during the
development of the showers.  However, in the present work we do not attempt to
correct for this effect. 

After a description of the experimental situation (\sref{measurements}), the
experimental results and comparisons with air shower simulations are described
in \sref{eas}.

\section{Experimental set-up}\label{measurements}

\subsection{The apparatus}

The experiment KASCADE, located on site of the Forschungszentrum Karlsruhe, 110
m a.s.l., consists of several detector systems. A description of the
performance of the experiment can be found elsewhere \cite{kascadenim}. A $200
\times 200$~m$^2$ array of 252 detector stations, equipped with scintillation
counters, measures the electromagnetic and, below a lead/iron shielding, the
muonic components of air showers. In its center, an iron sampling calorimeter
of $16 \times 20$~m$^2$ area detects hadronic particles. The calorimeter is
equipped with 11 000 warm-liquid ionization chambers arranged in nine layers.
Due to its fine segmentation ($25\times25$~cm$^2$), energy, position, and angle
of incidence can be measured for individual hadrons. A detailed description of
the calorimeter and its performance can be found in \cite{kalonim}, it has been
calibrated with a test beam at the SPS at CERN up to 350~GeV particle energy
\cite{kalocern}.

\subsection{Observables and event selection}

The position of the shower axis and the angle of incidence of a cascade are
reconstructed by the array detectors. The total numbers of electrons $N_e$ and
muons $N_\mu$ are determined by integrating their lateral distributions. In
case of muons, the ‘truncated muon number’ $N_\mu'$ is used for experimental
reasons. It is the number of muons integrated in the distance range $40-200$~m
from the shower axis. For a detailed description of the reconstruction
algorithms see \cite{kascadelateral}.
The position of the shower axis is reconstructed with an accuracy better than
2~m and the angle of incidence better than 0.5\deg.  

The hadrons in the calorimeter are reconstructed by a pattern recognition
algorithm, optimized to recognize as many hadrons in a shower core as possible.
Details can be found in \cite{kascadelateral}.  Hadrons of equal energy can
still be separated with a probability of 50\% at a distance of 40~cm. The
reconstruction efficiency rises from 70\% at 50~GeV to nearly 100\% at 100~GeV.
The energy resolution improves from 30\% at 50~GeV to 15\% at $10^4$~GeV.  The
hadron number $N_h$ and hadronic energy sum $\sum E_h$ are determined by the
sum over all hadrons in a distance up to 10~m from the shower axis. A
correction for the missing area beyond the boundaries of the calorimeter is
applied. 
The hadron lateral distributions are relatively steep, 
the hadronic energy density decreases by about two orders of magnitude
within the first 10~m from the shower
axis
 \cite{wwtestjpg}. Therefore, it is sufficient to measure hadrons in a
relatively narrow range around the shower axis only in order to collect a
significant fraction of the total hadron energy. 
The observable $\sum E_h$ includes also energy of
hadrons which could not be reconstructed independently, because they are too
close to each other. It shows up in the simulated and experimental data in the
same manner.

To be accepted for the present analysis, an air shower has to fulfill several
requirements: at least one hadron has been reconstructed in the calorimeter
with an energy larger than 50~GeV, the shower axis is located inside the
calorimeter, the electromagnetic shower size $N_e$ is larger than $10^4$, the
truncated muon number $N_\mu'$ is larger than $10^3$, i.e.\ the primary energy
is greater than about $3\cdot10^5$~GeV, and the reconstructed zenith angle is
smaller than 30\deg.
From May 1998 to October 2005 312000 showers have been measured meeting the
criteria mentioned.

To avoid corrections for different angles of incidence the following analysis
is restricted to showers with zenith angles $\Theta<18^\circ$.
The primary energy $E_0$ of the shower inducing particle is roughly estimated
based on the number of electrons $N_e$ and muons $N_\mu'$ registered with the
KASCADE scintillator array 
\begin{equation}\label{energyeq}
 \lg E_0 \approx 0.19 \lg N_e + 0.79 \lg N_\mu' + 2.33 .
\end{equation}
The average ground pressure during the observation time amounts to 1004~hPa,
corresponding to an average atmospheric column density $X_0$=1023~\gcm2. The
attenuation is measured at this depth $X_0$, the average vertical thickness of
the atmosphere above the KASCADE experiment.

\subsection{Simulations}

\begin{figure}[t] \centering
  \includegraphics [width=\columnwidth]{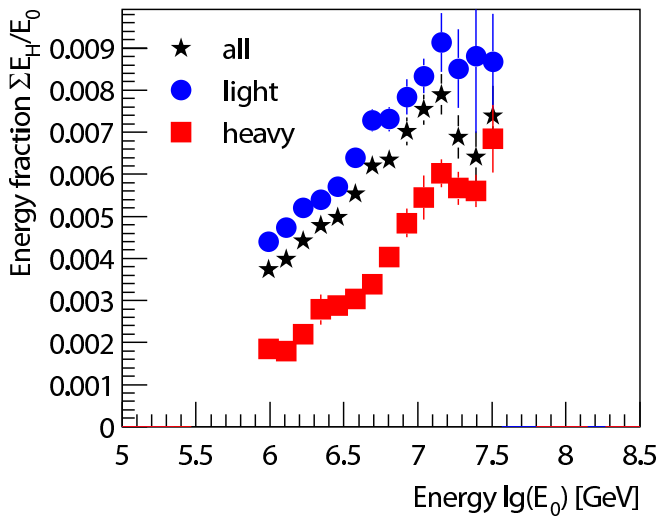}
  \caption{Fraction of energy $\Sigma E_H/E_0$ reaching the ground in form of
           hadrons as function of estimated primary energy $E_0$ for all data
	   and for selections of light and heavy primary particles.}
  \label{efrac}
\end{figure}

The shower simulations were performed using CORSIKA \cite{corsika}.  Hadronic
interactions at low energies were modeled using the FLUKA code
\cite{flukacern,flukaCHEN}.  High-energy interactions were treated with
QGSJET\,01 \cite{qgsjet} ($E>200$~GeV).  Showers initiated by primary protons
as well as helium, carbon, silicon, and iron nuclei have been simulated.  The
simulations covered the energy range $10^{5}-10^{8}$~GeV with zenith angles in
the interval $0^\circ - 32^\circ$.  The energy distribution of the showers
followed a power law with a spectral index of $-2.0$.  For the analysis the
energy distribution was converted to a power law with an index of $-2.7$ below
and $-3.1$ above the knee with a rigidity dependent knee position
($3\cdot10^6$~GeV for protons).  The positions of the shower axes are
distributed uniformly over an area exceeding the calorimeter surface by 2~m on
each side. In order to determine the signals in the individual detectors, all
secondary particles at ground level are passed through a detector simulation
program using the GEANT package \cite{geant}.  In this way, the instrumental
response is taken into account and the simulated events are analyzed by the
same code as the experimental data, an important aspect to avoid biases by
pattern recognition and reconstruction algorithms.  

\section{Results}\label{eas}

\subsection{Surviving Hadronic Energy}

The energy of the primary particle is estimated from measurements of the number
of electrons and muons in the shower with the scintillator array, see
\eref{energyeq}.  The surviving energy in form of hadrons $\Sigma E_H$ is
measured with the hadron calorimeter.  A fraction $R$, see \eref{fractioneq} of
hadronic energy reaching ground level can be inferred as function of primary
energy, as shown in \fref{efrac}.  All error bars represent statistical
uncertainties only.  Below $10^6$~GeV the values are affected by reconstruction
efficiencies. In particular, showers induced by heavy elements are less likely
to be registered.  Therefore, values are shown only for energies exceeding
$10^6$~GeV.  Above $10^7$~GeV the flux of the light cosmic-ray component
decreases and the composition becomes more and more heavy \cite{ulrichapp}.
Mos likely, this causes the structures seen in the figure for energies
exceeding $10^7$~GeV.  In the energy range investigated about 0.3\% to 0.8\% of
the primary energy reaches the observation level in form of hadrons, most of
them being pions \cite{mielkesh}.  

In the energy range of interest the elasticity of pions depends only weakly on
energy and can be approximated as $\epsilon\approx0.25$ to 0.3 \cite{wq}.  With
the relation $R=\epsilon^N$, the average number of generations $N$ in the
shower can be estimated and it turns out that the registered hadrons (with
energies above 50~GeV) have undergone about four to five interactions only.
This number is confirmed by full air shower simulations.

The fraction of hadronic energy reaching observation level increases with
energy, since the effect of deeper penetrating showers clearly dominates over
the small effect caused by the increase of the inelastic cross sections.

\begin{figure}[t] \centering
  \includegraphics [width=\columnwidth]{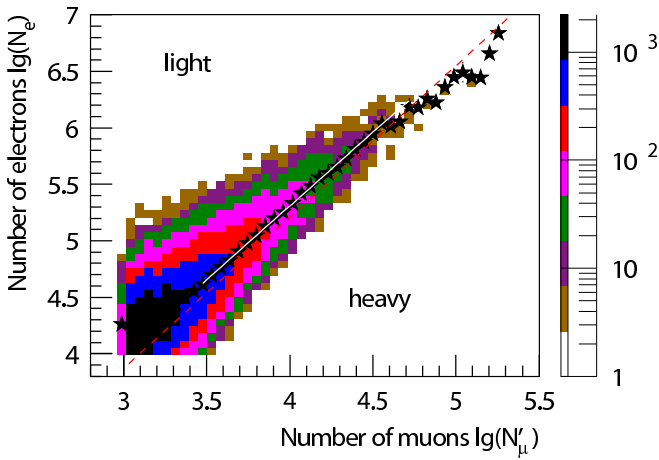}
  \caption{Number of electrons and muons for measured showers with zenith angle
           $\Theta<18^\circ$.
           The most probable values of the distribution are indicated by
	   the asterisks, the solid line represents a fit to this data.
           The dashed line represents \eref{cut}.}
  \label{comet}
\end{figure}

\begin{figure*}[t] \centering
  \includegraphics [width=\columnwidth]{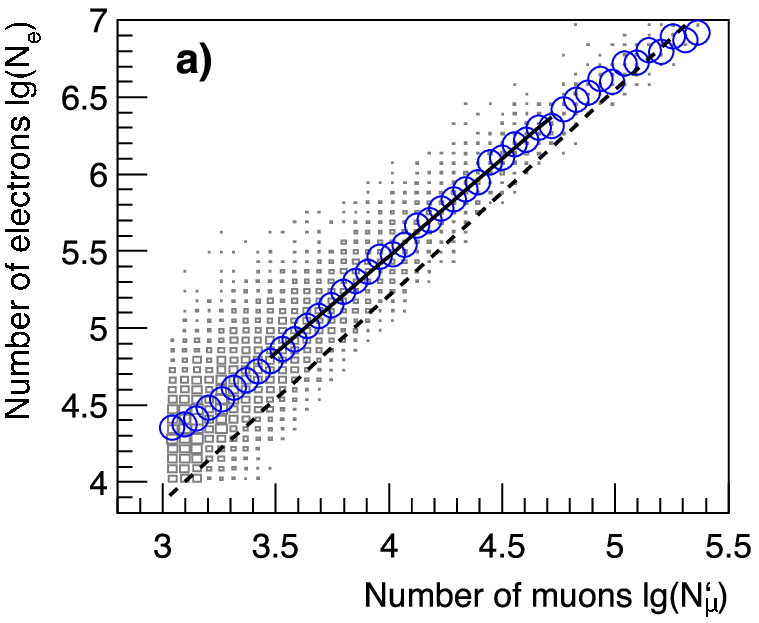}\hspace*{\fill}
  \includegraphics [width=\columnwidth]{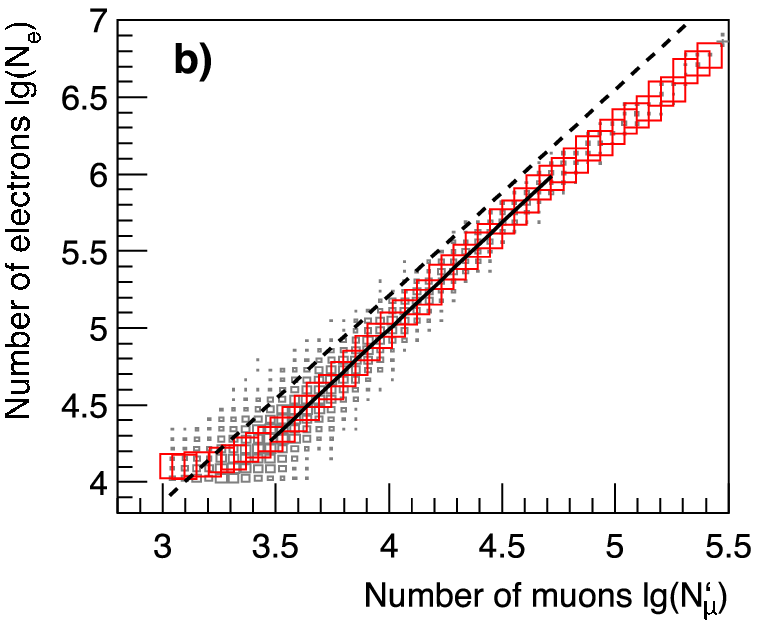}
  \caption{Number of electrons vs. number of muons in simulated air showers for
	   primary protons (a) and iron nuclei (b). The solid line
	   indicates a fit to the most probable values (circles and squares,
           respectively), the dashed lines represent \eref{cut}.}
 \label{comet-sim}	   
\end{figure*}

The two-dimensional distribution of the number of electrons and muons for the
measured showers is depicted in \fref{comet}. The asterisks represent the most
probable values of the distribution.  Also simulated $N_e-N_\mu'$ distributions
have been investigated for primary protons, as well as helium, carbon, silicon,
and iron-induced showers. Examples for protons and iron are depicted in
\fref{comet-sim}. The solid lines represent fits to the most probable values
represented by the circles and squares, respectively. It turned out that the
slopes of the fits of all elements are about equal. The fitted line for carbon,
parameterized as
\begin{equation}\label{cut}
 \lg(N_e)=1.34\lg(N_\mu')-0.15  
\end{equation}
is used in the following to divide the data into a sample induced by ``light''
and ``heavy'' primary particles, respectively.  It is indicated in
\ffref{comet} and \ref{comet-sim} as dashed lines.  In \fref{comet-sim} it can
be seen that \eref{cut} indeed separates the data set into ``light'' and
``heavy''. Almost no iron showers are above the dashed line and only a small
fraction of proton induced showers is below the line.  Especially the most
probable values for protons and iron-induced showers are clearly above and
below the dashed line, respectively. Parameterization \eref{cut} almost
coincides with the most probable values of the measurements.

Applying the selection criterion \eref{cut} to the data, the energy fraction
reaching observation level is shown in \fref{efrac} as well for light and heavy
primaries. As expected from a simple superposition model, proton-like showers
penetrate deeper into the atmosphere and transport more energy to the
observation level as compared to iron-like showers.

\subsection{Attenuation Length}

\begin{figure*}[t]
 \includegraphics[width=\columnwidth]{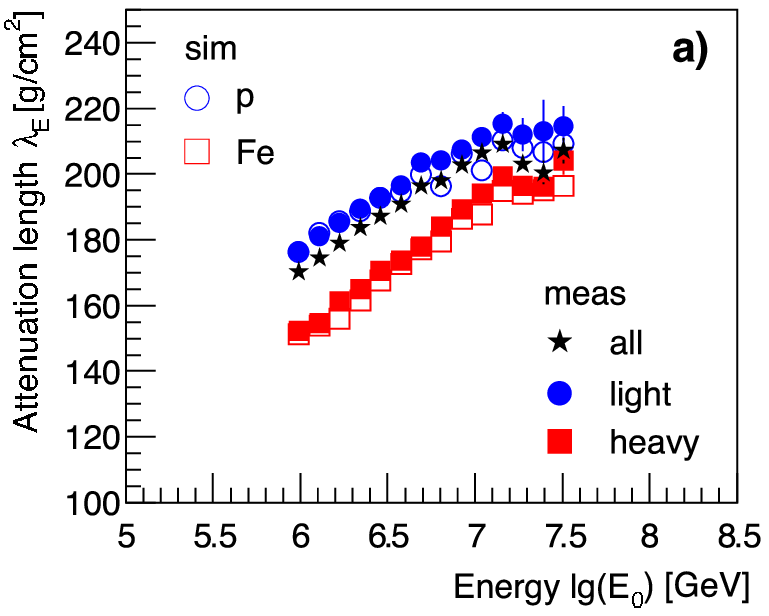}\hspace*{\fill}
 \includegraphics[width=\columnwidth]{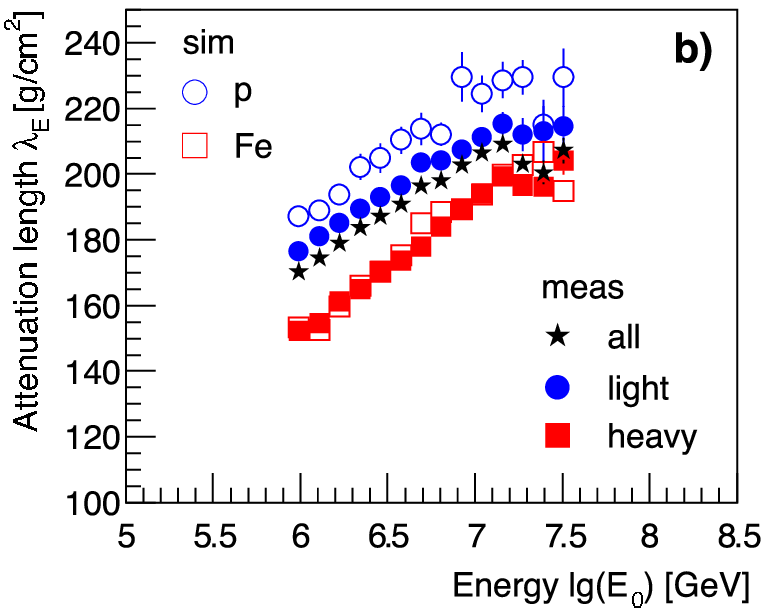}
 \caption{Attenuation length $\lambda_E$ as function of estimated primary
          energy.
	  The light and heavy groups in the measurements are compared to
	  simulations for primary protons and iron-induced showers using
          CORSIKA with the hadronic interaction model QGSJET\,01 (a)
	  and a modified version with lower cross sections and higher
          elasticity (b, model~3a in \rref{wq}).}
 \label{lep}
\end{figure*}

\begin{figure*}[t] \centering
  \includegraphics [width=\columnwidth]{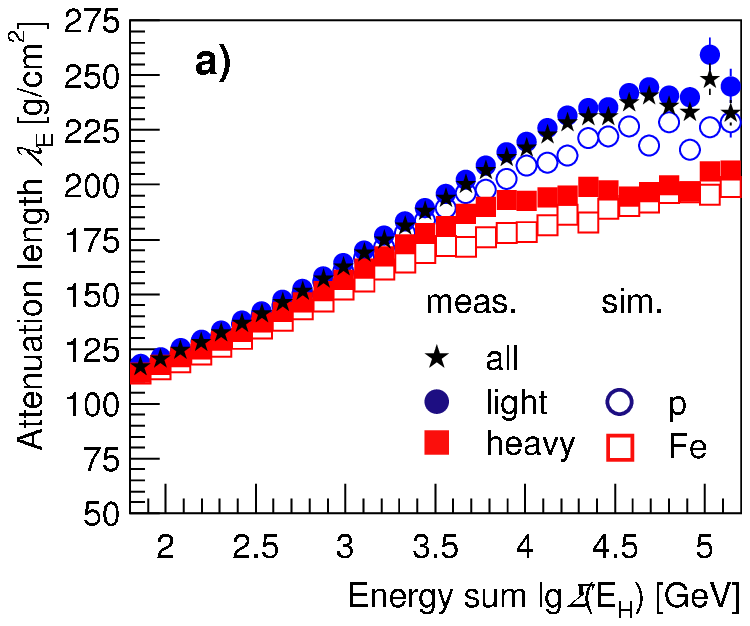}\hspace*{\fill}
  \includegraphics [width=\columnwidth]{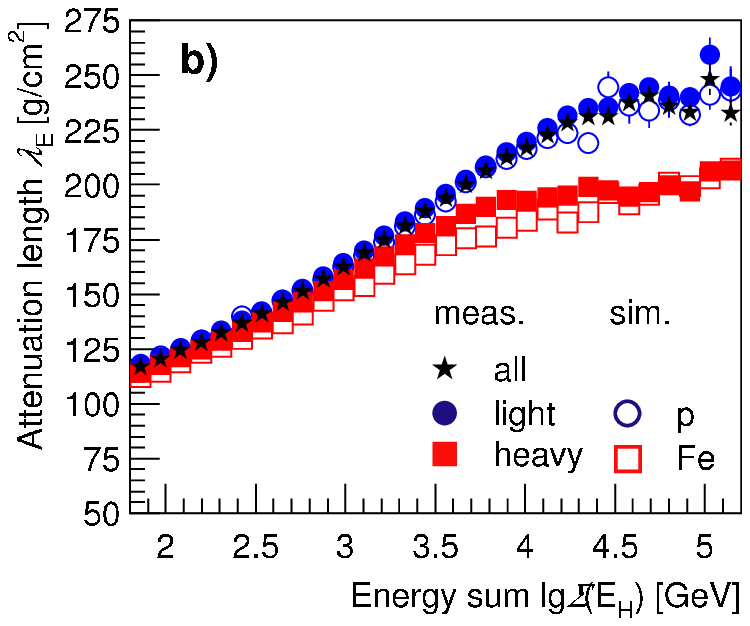}
  \caption{Attenuation length $\lambda_E$ as function of the measured
           hadronic energy sum at observation level.
	  The light and heavy groups in the measurements are compared to
	  simulations for primary protons and iron-induced showers using
          CORSIKA with the hadronic interaction model QGSJET\,01 (a)
   and a modified version with lower cross sections and higher
          elasticity (b, model~3a in \rref{wq}).}
  \label{leh}
\end{figure*}

Using the energy absorbed in the atmosphere at a depth $X_0$, the attenuation
length $\lambda_E$ has been derived from the measured energy fraction, see
\eref{ratioeq}.  The results are presented as function of the estimated primary
energy in \fref{lep}  a).  Values for the complete data set as well as for
the light and heavy selections are shown.  The values are compared to results
obtained from full air shower simulations for primary protons and iron nuclei
using the CORSIKA program with the hadronic interaction generator QGSJET\,01.
The cuts for a ``light'' and ``heavy'' component were applied to the proton and
iron simulations in the same way as for the measurements.  

The values shown in \fref{lep} have been obtained by applying the same
(quality) cuts and reconstruction algorithms to measured and simulated data.
Thus, many uncertainties are expected to cancel.  For the remaining differences
between simulated and measured data a systematic error for the hadronic energy
sum ($\sum E_H$) and the total energy ($E_0$) of 10\% each is assumed.  With
\eref{fractioneq} this results in a 14\% uncertainty of $R$.  In turn,
following \eref{ratioeq} yields a systematic error for $\lambda_E$ of order of
2\%. Therefore, the values shown in \fref{lep} have a systematic uncertainty of
order of 2 to 4~\gcm2. As discussed above, the structures above an energy of $10^7$~GeV are most likely due to the chaning composition as function of energy.

A comparison of the ``heavy'' selection with iron-induced showers shows that
the ``heavy'' selection lies slightly above the simulated values. This makes
sense, since the measured data contain a mixture of many elements, most of them
being lighter than iron, thus the measurements should be above the iron points.
On the other hand, looking at the ``light'' selection compared to the proton
points from the simulations, one recognizes that at high energies data points
are above the simulated values. This cannot be explained by a mixed composition
and may be a hint towards a problem in the hadronic interaction model
QGSJET\,01.  A possible explanation is that the attenuation length is too
small, i.e.\ the cross section is too large.

To test this hypothesis simulations have been carried out with a modified
version of QGSJET\,01, namely model~3a in \rref{wq}.  The inelastic hadronic
cross sections have been lowered, e.g.\ the proton-air cross section at
$10^6$~GeV is reduced by 5\% from 385~mb to 364~mb and the elasticity has been
increased by about 12\%. 
A similar trend to lower cross sections has been found as well by the EAS-TOP
experiment, with a value of 
$$\sigma^{inel}_{p-air}=338\pm21_{\mbox{(stat)}} \pm19_{\mbox{(sys)}}
-29_{\mbox{(sys-He)}}~\mbox{mb}$$ 
at $\sqrt{s}=2$~TeV ($\approx2\cdot10^6$~GeV) \cite{eastopwqprd}.
At the highest energies the lower proton-air cross section (443~mb at
$10^9$~GeV) is compatible with recent results from the HiRes experiment
$$\sigma^{inel}_{p-air}=456\pm17_{\mbox{(stat)}} +39_{\mbox{(sys)}}
-11_{\mbox{(sys)}}~\mbox{mb}$$ 
at $3\cdot10^9$~GeV \cite{belovisvhecri}.  
The lower cross sections have been proposed originally to reduce the
discrepancy in the mean logarithmic mass derived from experiments observing
shower maximum and investigating particle distributions at ground level
\cite{wq,isvhecri04wq}.  Applying the altered version of QGSJET also slightly
modifies the number of electrons and muons predicted at ground level.  At
energies around the knee ($\approx4\cdot10^6$~GeV) the number of electrons
increases by about 5\% and the number of muons rises by about 15\% \cite{wq}.  

The corresponding results for $\lambda_E$ are presented in \fref{lep} b).
The typical energies of particles in a shower induced by a heavy nuclei are
smaller than the typical energies in a proton-induced shower of the same
primary energy (superposition model of showers). Thus, the effect of the
modifications is stronger for proton-induced showers in \fref{lep} b), see
also \cite{wq}.  In contrast to \fref{lep}  a) the simulated points for
protons are now above the data points for the ``light'' selection. As a result,
with the modified version of the interaction model an overall improvement of
the situation has been achieved.

It should be pointed out that the experimentally accessible attenuation length
$\lambda_E$ is extremely sensitive to the inelastic hadronic cross sections. A
relatively small modification ($\sigma_{p-air}$ is changed by 5\% only at
$10^6$~GeV) yields significant changes, as can be inferred from \fref{lep}.

In addition, the results have been calculated also as function of the hadronic
energy sum at observation level.  The results for all data, as well as for the
light and heavy selections are presented in \fref{leh} a).  Again, a closer
inspection yields unreasonable results when the ``light'' elements are
considered. At high energies even all measured events are above the proton
simulations. In agreement with the previous discussion, the issue can be
resolved by introducing a modified version of the interaction model, as can be
seen in \fref{leh} b). Arranging the data in $\Sigma E_H$ bins implies an
enrichment of light primaries, which explains why all data almost agree with
pure proton simulations and the discrepancy between measurements and QGSJET~01
predictions are magnified.

\section{Summary and Conclusions}

A new method has been developed to derive the attenuation length of hadrons
from measurements of high-energy cosmic rays interacting in the Earth's
atmosphere.  The fraction of the energy of the primary particle reaching ground
level in form of hadrons in air showers has been measured with the KASCADE
experiment to increase from about 0.3\% at $10^6$~GeV to 0.8\% at
$3\cdot10^7$~GeV.  An attenuation length based on the absorbed energy has been
defined.  Corresponding values increase with energy from about 170~\gcm2 to
$\approx210$~\gcm2.

A closer inspection of the attenuation lengths obtained for showers induced by
``light'' and ``heavy'' elements indicates that the cross sections in the
hadronic interaction model QGSJET\,01 may be too large and the elasticity may
be too small. A modification with altered parameters improves the situation.

As final remark, it should be pointed out that the attenuation length
$\lambda_E$ is extremely sensitive to inelastic hadronic cross sections.  The
sensitivity of air shower measurements, in particular of the hadronic component
to properties of hadronic interactions has been demonstrated previously.  For
example, the dependence of observable quantities on the transverse momentum in
hadronic interactions \cite{annaprd} or on low-energy inelastic cross sections
\cite{epostest}.  In a similar way, the data presented and the method
introduced in the present article may serve to check and improve further
hadronic interaction models.

\begin{acknowledgments}
The authors would like to thank the members of the engineering and technical
staff of the KASCADE-Grande collaboration, who contributed to the success of
the experiment. The KASCADE-Grande experiment is supported by the BMBF of
Germany, the MIUR and INAF of Italy, the Polish Ministry of Science and Higher
Education together with the DAAD (PPP grant for 2009-2010), and the Romanian
Ministry of Education and Research (grant CEEX 05-D11-79/2005).
\end{acknowledgments}


\end{document}